\documentclass[review]{elsarticle}

\usepackage{hyperref}
\usepackage{lineno,hyperref}
\modulolinenumbers[5]

\usepackage{subcaption}
\usepackage{graphicx}
\usepackage{grffile}
\usepackage{comment}
\usepackage{tabularx}
\usepackage{multirow}

\hypersetup{colorlinks = true, pdfauthor=author, citecolor = blue}

\journal{Journal of \LaTeX\ Templates}

\usepackage{siunitx}

\begin{document}

\begin{frontmatter}

\title{Hypernuclear event detection in the nuclear emulsion with Monte Carlo simulation and machine learning}

\author[riken,gifu_eng,rikkyo]{A. Kasagi\corref{mycorrespondingauthor}}
\cortext[mycorrespondingauthor]{Corresponding author}
\ead{ayumi.kasagi@rikkyo.ac.jp}

\author[riken,saitama]{W. Dou}
\author[riken,gro]{V. Drozd}
\author[riken]{H. Ekawa}
\author[riken,csic]{S. Escrig}
\author[riken,imp,ucas]{Y. Gao}
\author[riken,lanzhou]{\\Y. He}
\author[riken,imp,ucas]{E. Liu}
\author[riken,gik]{A. Muneem}
\author[riken]{M. Nakagawa}
\author[riken,gifu_eng,gifu_fac_edu]{K. Nakazawa}
\author[csic]{\\C. Rappold}
\author[riken]{N. Saito}
\author[riken,lanzhou,gsi]{T.R. Saito}
\author[riken,saitama]{S. Sugimoto}
\author[rikkyo]{M. Taki}
\author[riken]{\\Y.K. Tanaka}
\author[riken,saitama]{A. Yanai}
\author[riken,tohoku]{J. Yoshida}
\author[nishina]{M. Yoshimoto}
\author[riken]{and H. Wang}


\address[riken]{High Energy Nuclear Physics Laboratory, Cluster for Pioneering Research, RIKEN, 2-1 Hirosawa, Wako, Saitama 351-0198, Japan}
\address[gifu_eng]{Graduate School of Engineering, Gifu University, 1-1 Yanagido, Gifu 501-1193, Japan}
\address[rikkyo]{Graduate School of Artificial Intelligence and Science, Rikkyo University, 3-34-1 Nishi Ikebukuro, Toshima-ku, Tokyo 171-8501, Japan}
\address[saitama]{Department of Physics, Saitama University, Saitama, 338-8570, Japan}
\address[gro]{Energy and Sustainability Research Institute Groningen, University of Groningen, Groningen, Netherlands}
\address[csic]{Instituto de Estructura de la Materia, CSIC, Madrid, Spain}
\address[imp]{Institute of Modern Physics, Chinese Academy of Sciences, 509 Nanchang Road, Lanzhou, 730000, Gansu Province, China}
\address[ucas]{University of Chinese Academy of Sciences, Beijing 100049, China}
\address[lanzhou]{School of Nuclear Science and Technology, Lanzhou University, 222 South Tianshui Road, Lanzhou, Gansu Province, 730000, China}
\address[gik]{Faculty of Engineering Sciences, Ghulam Ishaq Khan Institute of Engineering Sciences and Technology, Topi, 23640, KP, Pakistan}
\address[gifu_fac_edu]{Faculty of Education, Gifu University, 1-1 Yanagido, Gifu 501-1193, Japan}
\address[gsi]{GSI Helmholtz Centre for Heavy Ion Research, Planckstrasse 1, D-64291 Darmstadt, Germany}
\address[tohoku]{Department of physics, Tohoku University, Aramaki, Aoba-ku, Sendai 980-8578, Japan}
\address[nishina]{RIKEN Nishina Center, RIKEN, 2-1 Hirosawa, Wako, Saitama 351-0198, Japan}

\begin{abstract}
This study developed a novel method for detecting hypernuclear events recorded in nuclear emulsion sheets using machine learning techniques.
The artificial neural network-based object detection model was trained on surrogate images created through Monte Carlo simulations and image-style transformations using generative adversarial networks.
The performance of the proposed model was evaluated using $\alpha$-decay events obtained from the J-PARC E07 emulsion data.
The model achieved approximately twice the detection efficiency of conventional image processing and reduced the time spent on manual visual inspection by approximately 1/17.
The established method was successfully applied to the detection of hypernuclear events.
This approach is a state-of-the-art tool for discovering rare events recorded in nuclear emulsion sheets without any real data for training.
\end{abstract}

\begin{keyword}
Machine learning\sep Mask R-CNN\sep Geant4-simulation\sep GAN\sep Nuclear emulsion\sep Alpha-decay\sep Hypernucleus
\MSC[2010] 00-01\sep  99-00
\end{keyword}

\end{frontmatter}


\section{Introduction}
Nuclear emulsion a photographic sheet used to track charged-particles is a well-established detector in nuclear and particle physics research.
The trajectories of charged particles penetrating the photographic sheets were recorded as a series of metallic silver grains and observed as tracks using optical microscopy.
The achievable spatial resolution corresponds to sub-$\rm{\mu m}$.

In nuclear physics, nuclear emulsions have led to the discovery of hypernuclei \cite{First_hypernucleus}, subatomic systems that contain a minimum of one hyperon.
Hyperons are particles with three quarks; however, they contain a minimum of one strange quark.
Since the first observation of a hypernuclear event, they have been intensively studied for seven decades to understand the extended framework of nuclear force, the so-called baryon-baryon interactions, under flavored SU(3) symmetry \cite{Danysz1}.

The experimental approach for detecting hypernuclear events using nuclear emulsions is effective because it enables event-by-event analysis of the production and/or decay of hypernuclei to identify their nuclides.
Several emulsion experiments conducted from the 60's to the 70's with $K^{-}$ meson beams examined single-$\Lambda$ hypernuclei and provided fundamental data on hypernuclear physics \cite{BOHM, GAJEWSKI, JURIC}.
In the experiments, technical staff visually observed microscopic images and searched for hypernuclei, requiring considerable effort.
The counter hybrid emulsion method was introduced in the 1980s to limit the search volume by providing supportive information from real-time detectors \cite{E176_2009}.
This method has been successful in searching for double-strangeness hypernuclei \cite{E176_2009, Nagara, Ahn}.
In the latest hypernucleus search experiment with nuclear emulsion, J-PARC E07, part of the operation of the microscope was automated \cite{E07_proposal, MYINTKYAWSOE201766}.
These improvements in efficiency resulted in the detection of three times as many candidates as in previous experiments \cite{MINO, IBUKI, Yoshimoto}.

Despite this success, the potential for further hypernuclear research lies in data from nuclear emulsion detectors.
The counter-hybrid method can only search for reactions that are set to trigger conditions using a real-time detector.
Considering the number of $K^{-}$ beams irradiated onto the emulsion sheets and the number of emulsion sheets of E07, several million single-$\Lambda$ hypernuclear events via direct reactions of $K^{-}$ beams inside the emulsion sheets can be detected.
If hypernuclear events can be detected automatically from microscopic images, the types of hypernuclei and number of events that can be analyzed will significantly increase.

\begin{figure}[htbp]
    \centering
    \includegraphics[width=120mm]{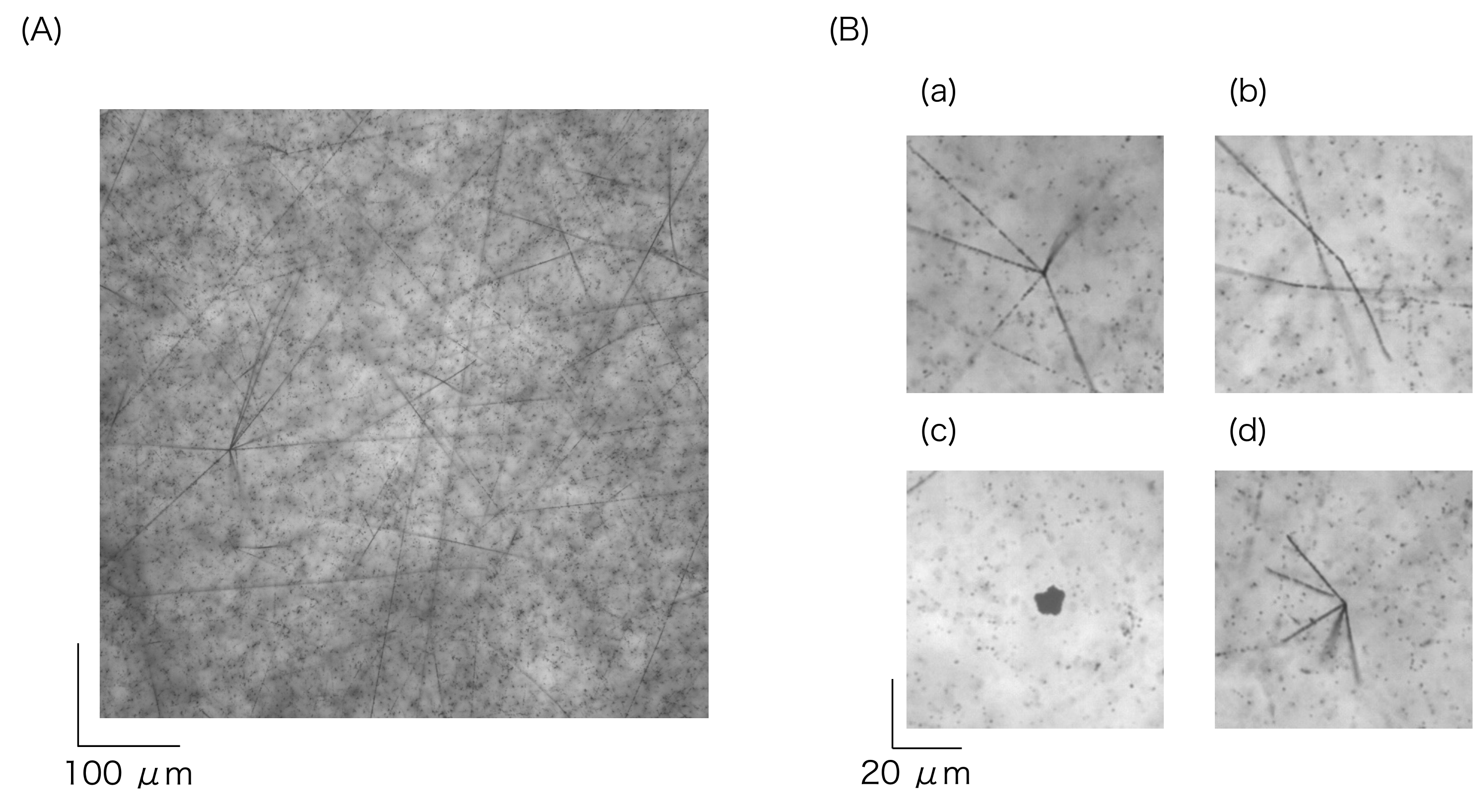}
    \caption{Typical microscopic images of a nuclear emulsion sheet for hypernuclear physics experiment and detected events.
    Panel (A) shows an example of single field-of-view images captured with an optical microscope. 
    Black lines and dots represent the trajectories of charged particles. 
    Panel (B) shows examples of detected events with the vertex picker, a conventional line detection-based method \cite{Yoshida_overall} 
    Sub-panels (a) and (b) show a $K^{-}$ beam interaction event and an unrelated track crossing event, respectively.
    Sub-panel (c) shows a shadow created by dust and/or other emulsion layers.
    Sub-panel (d) represents an $\alpha$-decay event, a decay chain event of a natural radioisotope $^{228}\rm{Th}$ to $^{208}\rm{Pb}$.
    \label{fig:emulsion_image}}
\end{figure}

Recent computer vision technologies have enabled the development of a detection method called overall scanning \cite{Yoshida_overall, Yoshimoto_hyp}.
The scanning technology of microscope systems has been able to read-out the entire volume of nuclear emulsion sheets within a few years.
Therefore, a novel event detector that can achieve high purity and efficiency is necessary to realize this method because the data size is large and hypernuclear events are rare.
Panel (A) of Figure~\ref{fig:emulsion_image} shows examples of microscopic images of a nuclear emulsion sheet.
The black lines and dots on the gray background represent tracks of the charged particles.
Approximately $10^6$ $K^{-}$ beam tracks per $\rm{cm}^{2}$ penetrate perpendicular to the focal plane, and many background tracks such as fragments of nuclear fragmentation reactions and cosmic rays accumulate without any time information.

To search for hypernuclear events in such images, Vertex picker, which detects vertices formed by particle tracks associated with hypernuclear production and/or decay, was developed based on line segment detection \cite{Yoshida_overall}.
Although the presence of $\Xi$ hypernuclear event was confirmed among 8 million images \cite{KISO}, most of the detected events were background events unrelated to hypernuclear events.
Panel (B) of figure~\ref{fig:emulsion_image} shows examples of the events detected by the vertex picker.
Sub-panels (a), (b), and (c) show a nuclear fragmentation reaction caused by $K^{-}$ beam and the nucleus in the emulsion sheet, a crossing event of unrelated tracks, and a shadow of a dust-like object in the emulsion layer.
In addition, as shown in sub-panel (d), $\alpha$-decay events, decay chains of natural radioisotopes in the raw material of the emulsion were also detected.
The number of objects detected by the vertex picker was approximately $10^{10}$ if the entire data of E07 were analyzed, and it would take over $5 \times 10^{2}$ years to classify by visual inspections even if ten people were involved \cite{Saito_nature}.
Therefore, to reduce analysis time, a new method was required.

In a previous study, an image classification method based on convolutional neural networks (CNNs) was employed \cite{Yoshida_cnn}.
The images of the events detected by the vertex picker were manually classified into $\alpha$-decay and other events.
Approximately 50,000 training datasets were prepared.
The trained model reduced the number of visual inspections by a factor of 1/7.
However, a challenge arose in obtaining training data from actual image data for hypernuclear events, unlike $\alpha$-decay events.

We propose and develop a novel method for detecting hypernuclear events in nuclear emulsions using Monte Carlo simulations and machine learning-based object detection methods.
The combination of these techniques addressed the inability to use actual image data as training data in hypernuclear event searches and achieved high detection performance.
 
The rest of this paper is organized as follows. Section 2 presents the concepts and development procedures.
The procedure for creating the training data is presented in Section 3.
Section 4 describes the development of an object detection model based on the Mask R-CNN.
Section 5 discusses the results and performance of the proposed method.

\clearpage
\section{Overview of the proposed method}
The proposed method employs an artificial neural network-based model, the Mask R-CNN \cite{Mask_RCNN}, which is a well-established object-detection model with various known applications. 
It is versatile and feasible to develop a detection model for a specific object of interest by preparing images containing the object as training data.
To train and evaluate the model, many images of training data for object detection are required. 
For a hypernucleus search, it is infeasible to obtain sufficient images.

This study overcomes that difficulty by applying an image-style transformation based on machine learning techniques.
These techniques convert the coordinate information of the trajectory of a charged particle, represented as a line segment, into a microscopic image.
Even for rare events, track information can be prepared by Geant4 \cite{Geant4}, which is a physics simulation framework using Monte Carlo methods.
The images employed for training the Mask R-CNN model were generated using an image-style translation model, pix2pix \cite{pix2pixHD} with generative adversarial networks (GANs) \cite{GAN}.
These produced images surrogate the real training data, which are typically used in machine learning applications.

The development and evaluation of this method were first conducted using $\alpha$-decay events before their application to hypernuclear events because they are moderately recorded in nuclear emulsions.
Furthermore, the detection of $\alpha$-decay is essential, because the emitted $\alpha$ particles are the calibration source for the range energy relation in nuclear emulsions \cite{Barkas, Barkas_2, Lin}.

\clearpage
\section{Creating training data}
The creation of training data consisted of three main steps: event generation by Monte Carlo simulation, compounding of background events, and conversion to a microscopic-like image using a style transformation technique.

\subsection{Generating events with Geant4 Monte Carlo simulations}
\begin{figure}[htbp]
    \centering
    \includegraphics[width=120mm]{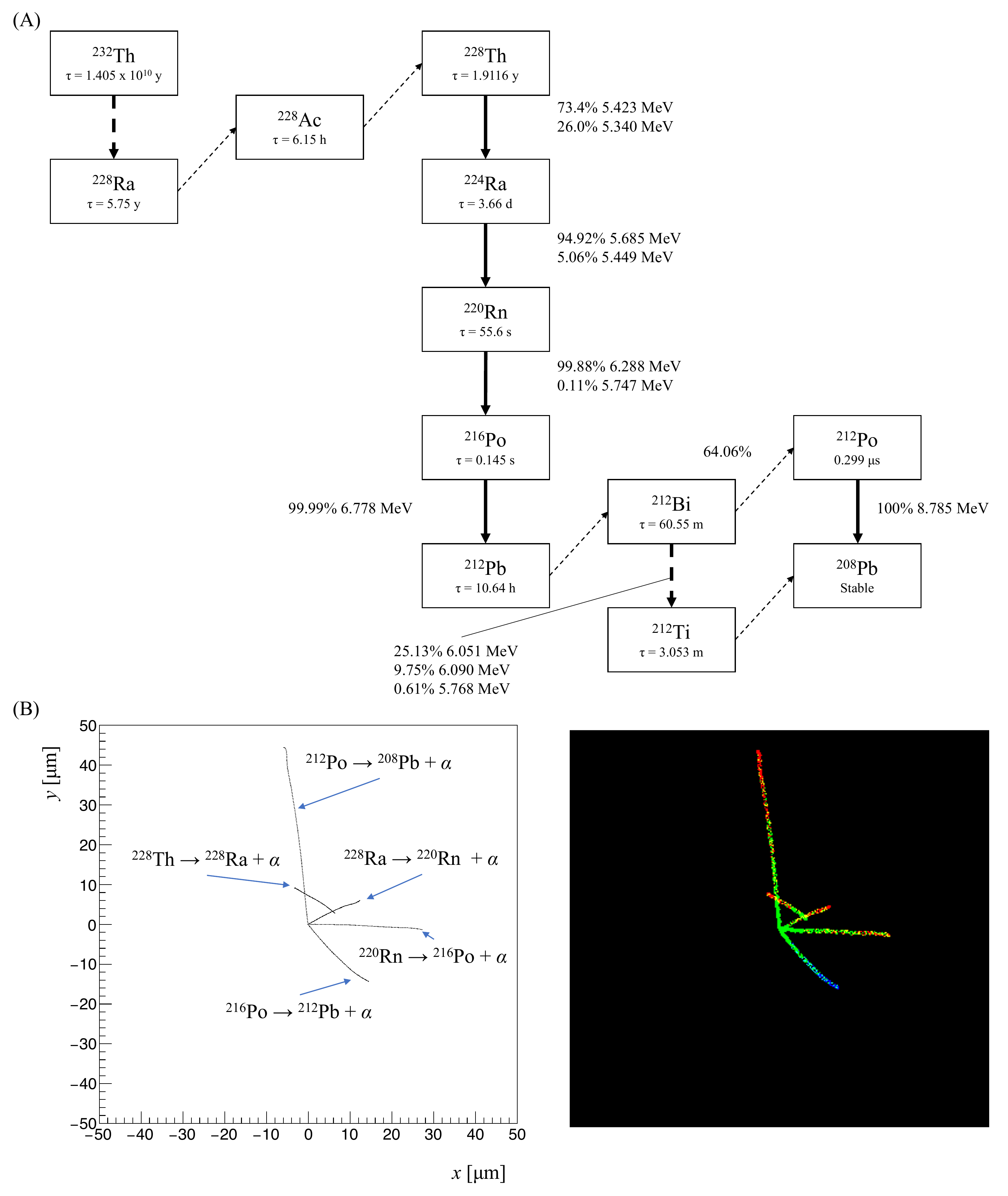}
    \caption{Chart of the decay chain of the thorium series and line images of simulated $\alpha$-decay events with Geant4.
    Panel (A) shows decay chain from $^{232}\rm{Th}$ to $^{208}\rm{Pb}$ with their half-life and energies of emitted $\alpha$ particles \cite{N208, N212, N216, N220, N224, N228}.
    The figure on the left of Panel (B) shows the trajectories of $\alpha$-particles corresponding to solid lines appearing in Panel (A).
    The trajectory information is converted to an image with an RGB channel to reproduce three-dimensional coordinates, as shown on the right side of Panel (B).
    The image has three layers, with the green layer representing the optimal focus plane and the red and blue layers representing the tracks in shallower and deeper planes, respectively.
    \label{fig:alpha_geant}}
\end{figure}

The geometry and elemental composition shown in Table \ref{tab:emulsion_composition} of a nuclear emulsion detector were reproduced in the Geant4 framework. 
Tracks associated to events were generated inside the detector to record the coordinates and energy deposition information of the particles.
Figure~\ref{fig:alpha_geant} shows the decay chain and an example of $\alpha$-decay events of the thorium series generated in the simulation.
$\alpha$ particles, represented by the solid lines in Panel (A)
appear as trajectories in the left figure of Panel (B).
A pixel size of $0.29~\rm{\mu m/pix}$ and a focus depth of $3~\rm{\mu m}$ were selected when the coordinates of the charged particles were visualized. These parameters were based on the capability of the microscope and the objective lens used to acquire images of the actual nuclear emulsion sheets.
RGB colors were used to represent the depth of the focal plane and provide three-dimensional information in the image; green represents tracks in the focal plane, and red and blue represent tracks in shallower and deeper planes, respectively.
Tracks across multiple layers are represented by composite colors that indicate the boundary between the in-focus and out-of-focus areas.

The apparent boldness of tracks in nuclear emulsions varies with their energy loss and zenith angle \cite{Kinbara}.
The energy loss of a particle passing through a nuclear emulsion is correlated with the number of grains generated in a unit length along the trajectory, which is the grain density.
The grain density was reproduced using the energy loss per recording step in the Geant4 simulation.
Furthermore, referring to the results of \cite{Kinbara}, who discussed a charge identification method by evaluating the thickness of the tracks corresponding to various nuclides, momentum, and zenith angles, the thickness was calculated based on the angle of each track and the momentum at each step.

\begin{table}[htbp]
    \centering
    \caption{Composition of E07 experimental emulsion layer at beam exposure which was estimated from the production process\label{tab:emulsion_composition}}
    \begin{tabular}{c|c|c|c|c|c|c|c}
        &H&C&N&O&Br&Ag&I\\
        \hline
        \hline
        Mass ratio [\%]&1.42&9.27&3.13&6.54&33.17&45.52&0.94\\
        Atomic ratio [\%]&38.56&21.10&6.10&11.17&11.34&11.53&0.20\\
    \end{tabular}
\end{table}

One effective technique for training machine learning models that achieves excellent classification and detection performance is to include a negative sample \cite{negative_sample}.
By including the predicted background events in the training data in advance, the parameters are expected to be updated to suppress false positives.
The dominant background events for detecting $\alpha$-decay events are the fragmentation reaction events shown in sub-panel (a) of Figure~\ref{fig:emulsion_image}.
They have identical characteristics to $\alpha$-decay events in that they are composed of multiple charged particle tracks emitted from the reaction point.

\begin{figure}[htbp]
    \centering
    \includegraphics[width=120mm]{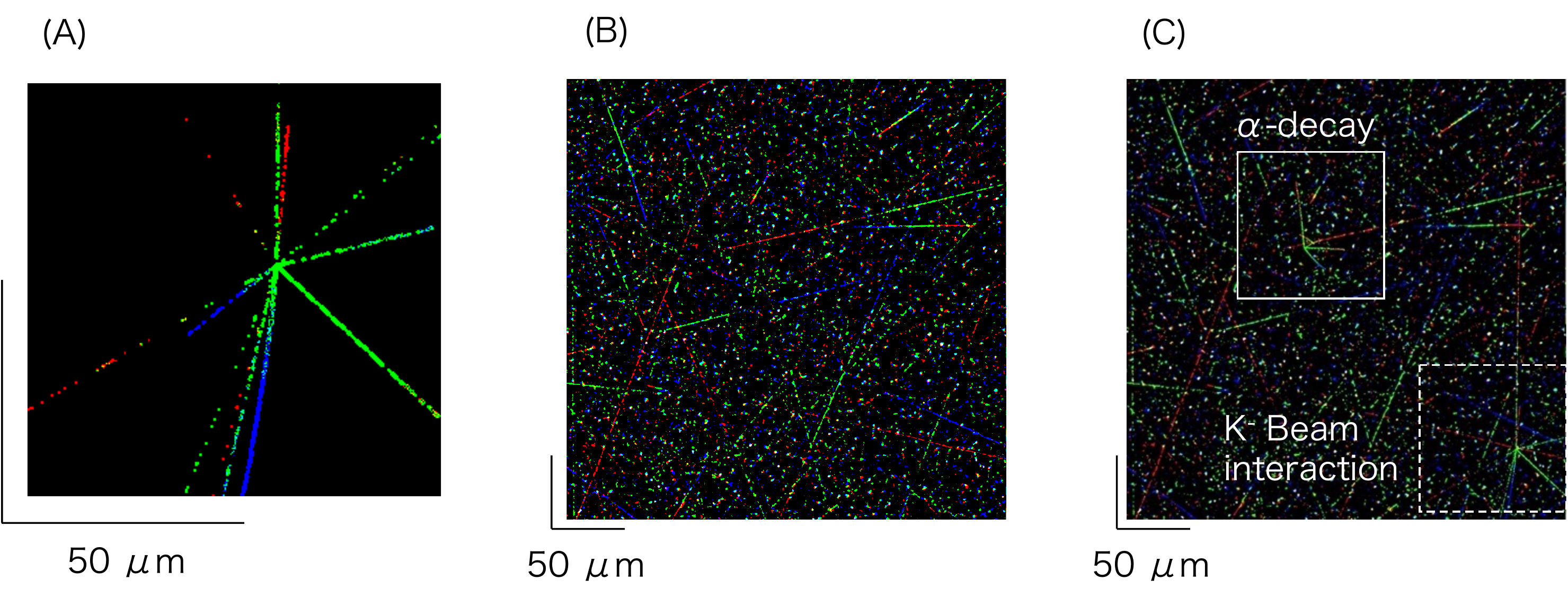}
    \caption{Signal and background events in colored images.
    Panel (A) shows a beam interaction event with associated tracks simulated by JAM \cite{JAM} and Geant4 simulation \cite{Geant4}
    Panel (B) shows an example of the background images converted from an actual microscopic image. 
    The signal event, as shown in Figure~\ref{fig:alpha_geant}, Panels (A) and (B) are mixed, and the colored image was produced as shown in Panel (C).
    \label{fig:mixed_images}}
\end{figure}

The JAM package \cite{JAM}, which is based on data from hadron scattering experiments, was used to simulate the tracks caused by fragmentation reactions.
The emitted particles were considered from the reaction of $K^{-}$ beam at 1.8~GeV/c bombarding medium and heavy nuclei in the gelatin of the emulsion layer, such as C, N, O, Ag, and Br.
Panel (A) in Figure~\ref{fig:mixed_images} shows an example of images of $K^{-}$ beam interaction event generated by JAM.
The tracks of the emitted charged particles were visualized using the same Geant4 framework that produced the $\alpha$-decay images.
High-momentum particles are represented by dotted lines owing to their low energy loss.

Other random background tracks and noise were obtained from actual microscope images.
Panel (B) in Figure~\ref{fig:mixed_images} shows the processed image of an actual microscopic image with a high-pass filter and binarization.
The tracks and noise recorded in the different layers were superimposed, and the RGB channel indicated their depth information in the same manner as the simulated events.

The training image was generated by integrating the event to be detected and the tracks that served as background events in the detection.
Panel (C) in Figure~\ref{fig:mixed_images} shows an image of the tracks associated with signal $\alpha$-decay events; tracks of background events are also added.
$\alpha$-decay and beam interaction events are shown as white rectangles with solid and dashed lines in the image, respectively.

\subsection{Image style transformation using pix2pix}
For the object detection model to provide information on the shape of an event, it is desirable to reduce differences in the training data.
Therefore, style transformation using GANs was introduced to transform image styles \cite{GAN}.
GANs are deep learning models used to generate new and synthetic data resembling an input dataset. 
They comprise two neural networks, a generator, and discriminator, which work simultaneously to generate the data. 
The generator network produces new samples, while the discriminator network attempts to determine the authenticity of each sample. 
During training, both networks were updated based on each other's performances. 
This process continues until the generator generates samples that are indistinguishable from the real data.

In this study, we adopt Pix2pix \cite{pix2pixHD} using a GAN, which is one of the implementations that have significantly improved the performance of image-style transformation through the adoption of U-NET \cite{UNET}.
A dedicated model was developed to transform line segments into microscopic images using this network.

\begin{figure}[htbp]
    \centering
    \includegraphics[width=120mm]{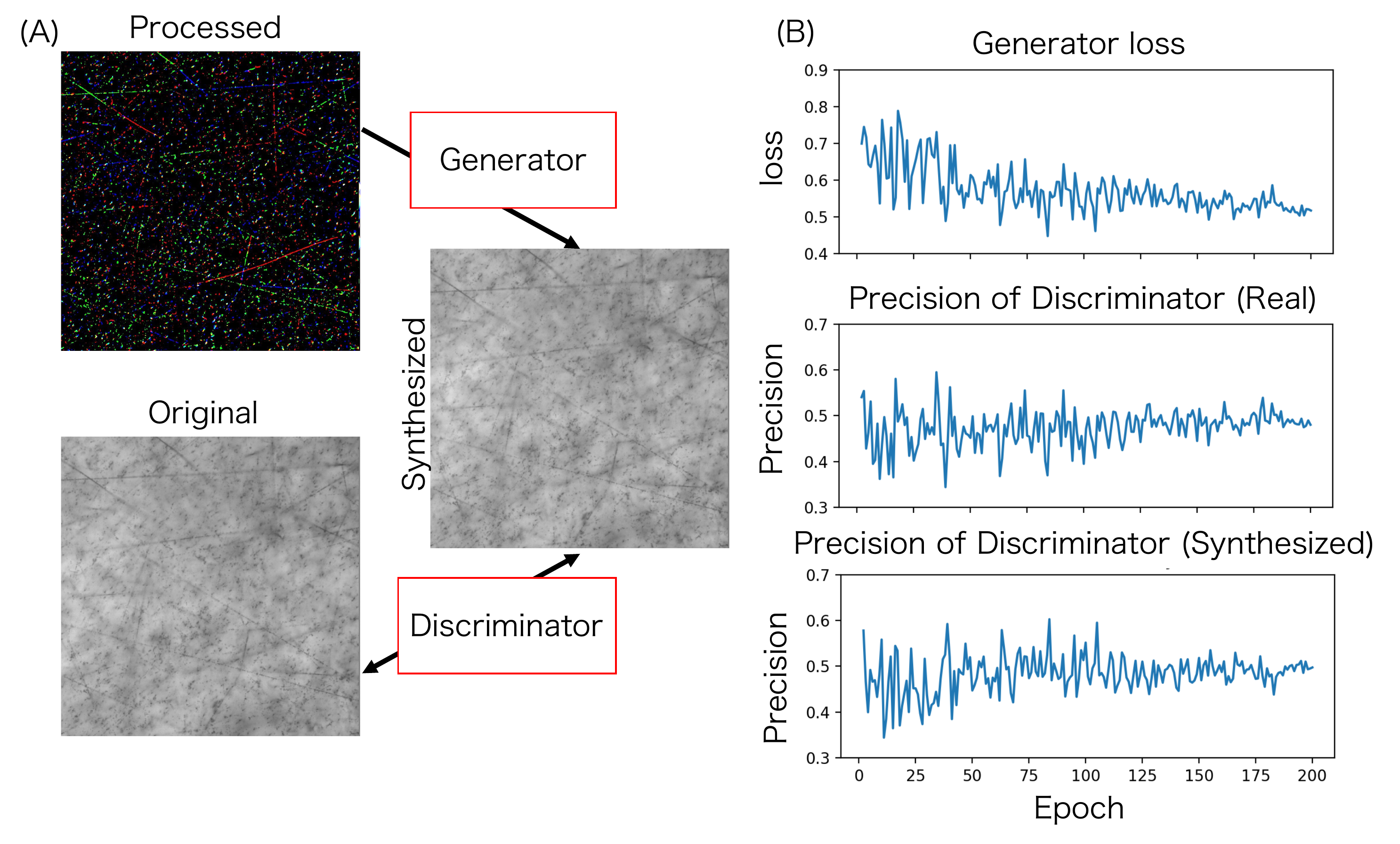}
    \caption{Panel (A) shows an example of images of training data for the pix2pix model and a generated synthesized image.
    The generator converts processed images to synthesized images similar to real microscopic images of nuclear emulsions. The discriminator attempts to distinguish and classify real and synthesized images.
    Panel (B) shows the learning curves corresponding to the performances of networks at each epoch.
     The generator's loss, the discriminator's classification precision for actual images, and the precision for synthesized images are represented by the top, middle, and bottom figures, respectively.
    \label{fig:pix2pix_training}}
\end{figure}

Panel (A) of Figure~\ref{fig:pix2pix_training} shows an example of the images used for pix2pix training data.
The image processing applied to the microscope images was the same as that used for background image production, as shown in Figure~\ref{fig:mixed_images} (B).
The model was trained using twenty thousand image pairs.
The learning curves are presented in Panel (B), and the hyperparameters for the training are summarized in Table \ref{tab:params_pix2pix}.
All other parameters not mentioned are in accordance with the settings of \cite{pix2pixHD}.

\begin{figure}[htbp]
    \centering
    \includegraphics[width=120mm]{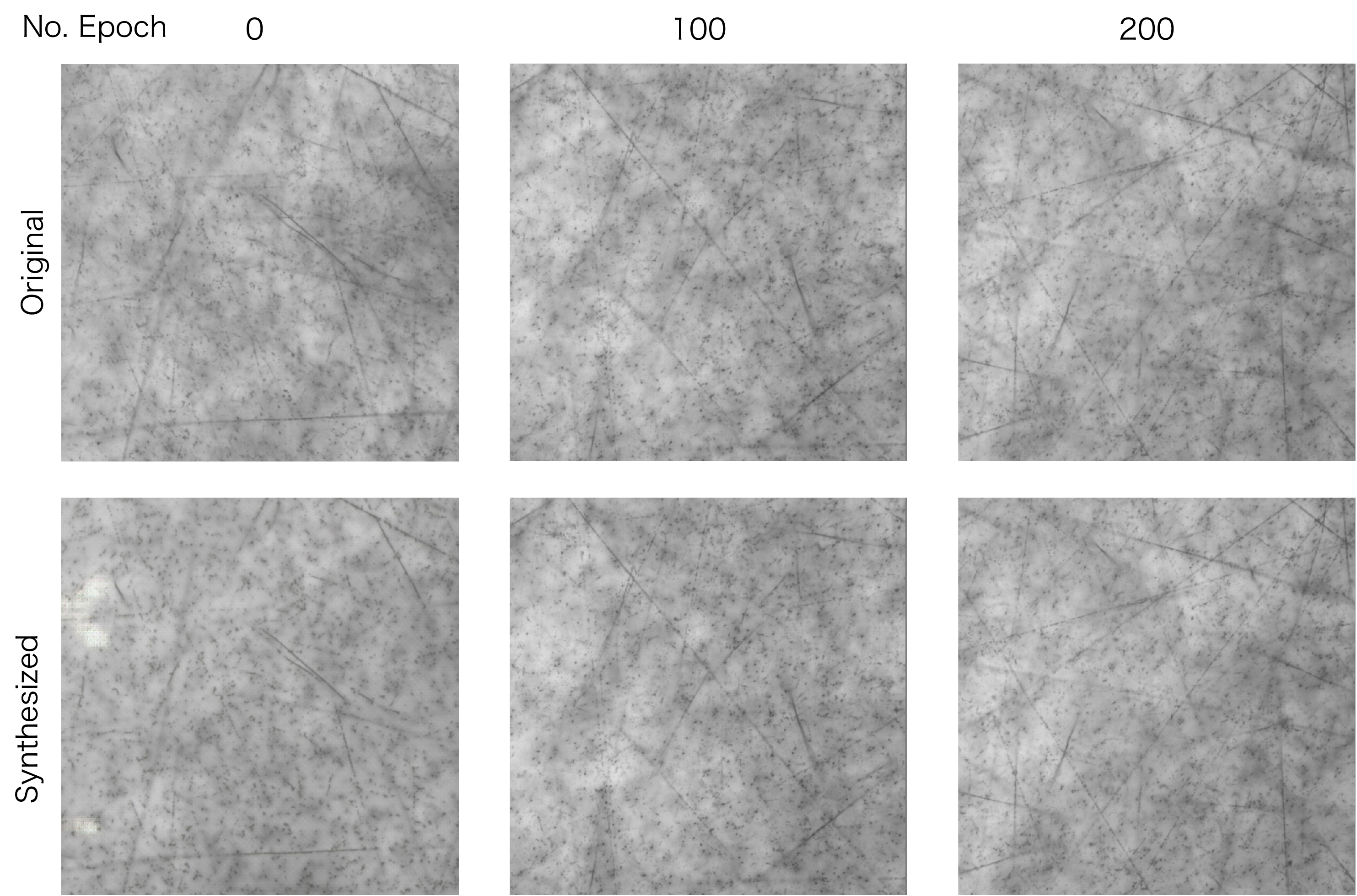}
    \caption{Comparisons of original and output images by the number of iterations.
    Numbers on top of the images correspond to counts of iteration.
    Based on these results and the trend of the training curves shown in Figure~\ref{fig:pix2pix_training}, the final parameters were used for subsequent style transfer. \label{fig:pix2pix_iteration}}
\end{figure}

Learning curves represent a metric of the performance of each network, corresponding to the number of epochs.
The generator loss corresponds to the value that quantifies the difference between the input data and the data synthesized by the generator, and the precision corresponds to the discriminator's classification success rate for real and synthesized images \cite{pix2pixHD}.
Comparisons between the original microscope and output images at different epochs are shown in Figure~\ref{fig:pix2pix_iteration}.
As the number of epochs increases, the trajectories, background noise, and defocusing images are reproduced more practically.
The parameters obtained at the 200th epoch are used for subsequent style conversion because the generator training loss and discriminator precision almost converged.

\begin{table}[htbp]
    \centering
    \caption{Hyperparameters for the training of pix2pix model. This development was performed based on the package of pix2pixHD implemented by PyTorch \cite{pytorch}\label{tab:params_pix2pix}}
    \begin{tabular}{c|c}
    Model&pix2pixHD \cite{pix2pixHD}\\
    \hline
    Batch size&1\\
    \hline
    Image size& $1024 \times 1024$ pixel\\
    \hline
    \hline
    Generator&\\
    \hline
    Network of generator&Local\\
    \hline
    Number of local enhancers&3\\
    \hline
    Number of residual blocks&3\\
    \hline
    \hline
    Discriminator&\\
    \hline
    Momentum term of adam \cite{adam} &0.5\\
    \hline
    Initial learning rate&0.0002\\
    \hline
    Number of discriminators&3\\
    \hline
    The number of the channels of first convolutional layer&64\\
    \hline
    Weight for feature matching loss&10.0\\
    \hline
    \hline
    Total epochs&200\\
    \end{tabular}
\end{table}

\begin{figure}[htbp]
    \centering
    \includegraphics[width=120mm]{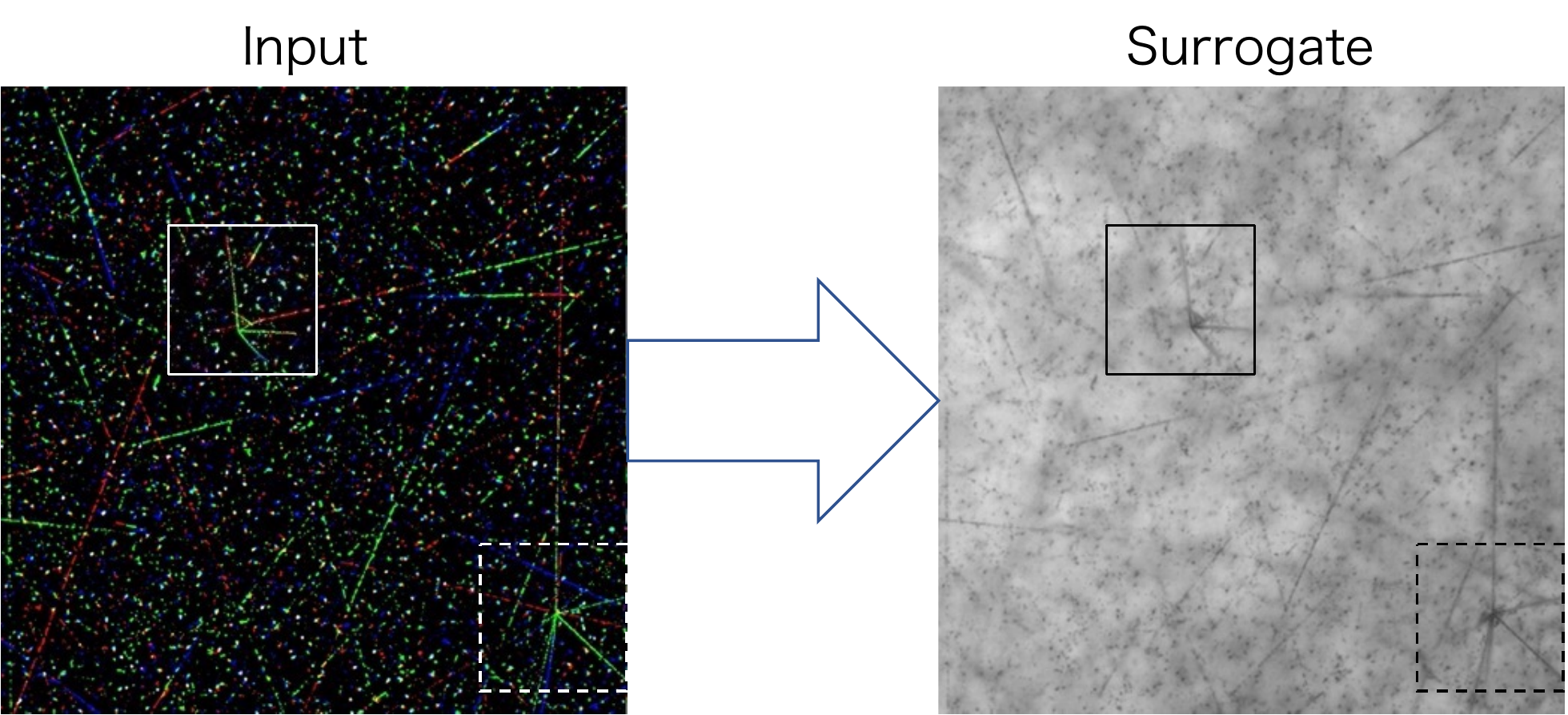}
    \caption{The input colored image is the same as shown in panel (C) of Figure~\ref{fig:mixed_images}. The microscopic-like image, including a simulated $\alpha$-decay event, a beam interaction, and other background events.
    $\alpha$-decay and beam interaction events in both images are shown as rectangles with solid and dashed lines, respectively.
    \label{fig:alpha_surrogated}}
\end{figure}

Figure~\ref{fig:alpha_surrogated} presents an example of the transformation of a simulated image using the trained pix2pix model.
The resulting images were applied as training data for the object detection machine learning model described in the subsequent section, which was a surrogate for real training data.

\clearpage
\section{Event detector based on Mask R-CNN}
We developed an object detection method based on a machine learning model called a Mask R-CNN \cite{Mask_RCNN}.
The adopted Mask R-CNN is a well-established object-detection model with various applications and implementations.
The functions for detecting the location of the events of interest in the input images and categorizing them were implemented in a single network.
Classification metrics were assigned scores ranging from zero to one.
In addition, the implemented segmentation features allowed us to obtain detailed shapes of the objects detected in the image.

\begin{figure}[htbp]
    \centering
    \includegraphics[width=120mm]{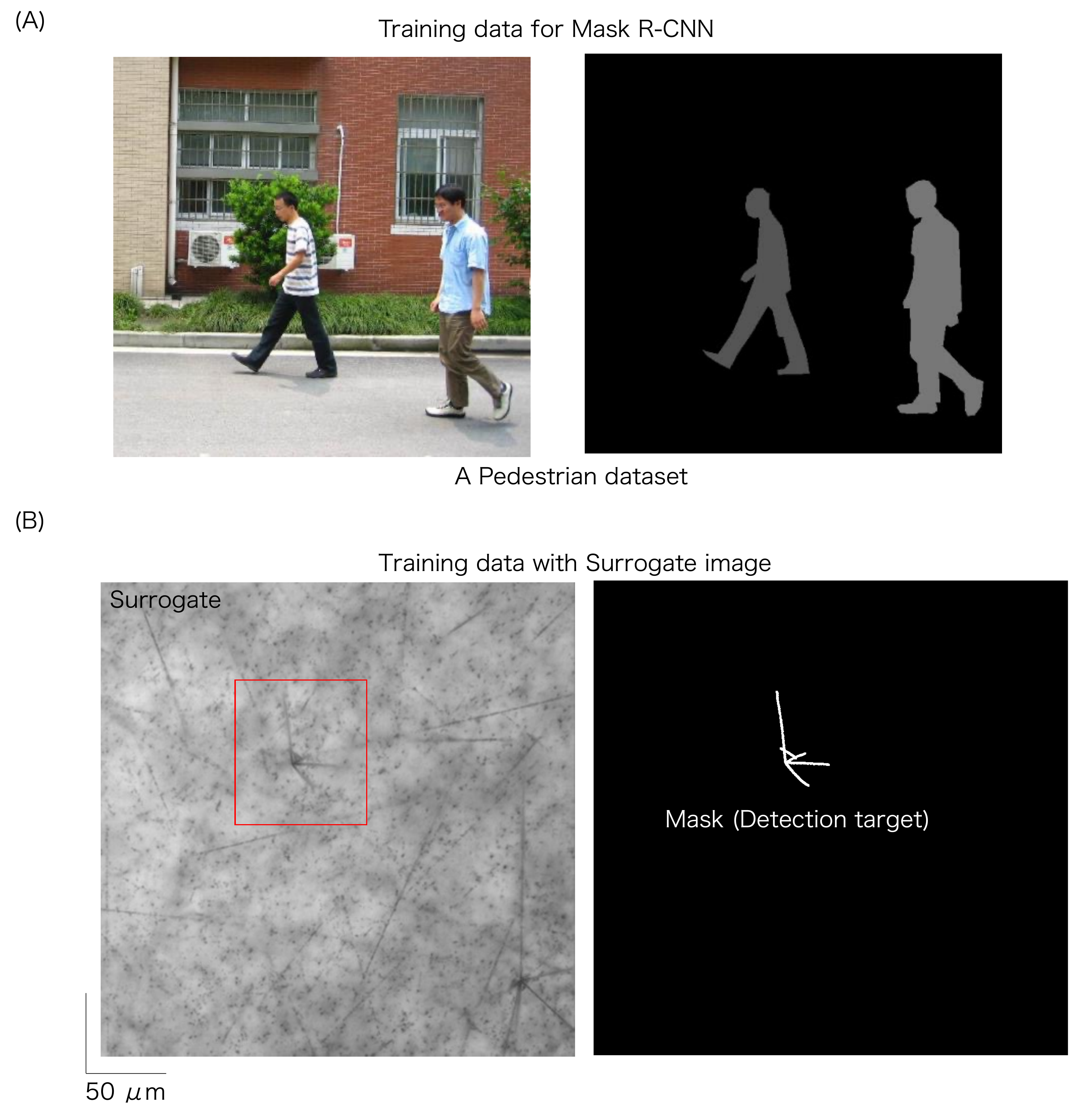}
    \caption{Examples of training image data for Mask R-CNN.
    Panel (A) shows a typical example of image pairs that consist of the input image (left) and mask information (right).
    The figures are obtained from the Penn--Fudan database \cite{Pedestrian}.
    The mask for pedestrians in the image was produced with manual annotation and represents the location and shape of the object.
    Panel (B) shows a pair of the image in this study generated by pix2pix and the mask image with the trajectory information associated with the event to be detected.
    Because the information on the trajectory has already been obtained in the simulation, the masks can be generated without manual annotation.
    \label{fig:maskrcnn_training}}
\end{figure}

The training data for the Mask R-CNN-based model consisted of the input and mask images containing information on the shape and location of the objective event.
Panel (A) of Figure~\ref{fig:maskrcnn_training} shows an example of a training dataset for pedestrian detection from the Penn--Fudan database \cite{Pedestrian}. 

Here, the pedestrians in the image were the objectives; hence, the location of a human in the mask image was represented by pixels with unique brightness values.
This mask image is typically created by manual work called ``annotation." 
This process is one of the most time-consuming tasks when creating an object detection model using machine learning, particularly when needing to prepare a large amount of training data.
In contrast, in our training data shown in Panel (B) of Figure~\ref{fig:maskrcnn_training}, the mask information given for the shape and position of the tracks composing the event can be simultaneously obtained by the Geant4 simulation.
Consequently, a large amount of training data was quickly prepared without annotation work in our development.
Thirty-thousand surrogate images and the corresponding mask information were used to develop a Mask R-CNN-based model for $\alpha$-decay detection.

\begin{figure}[htbp]
    \centering
    \includegraphics[width=120mm]{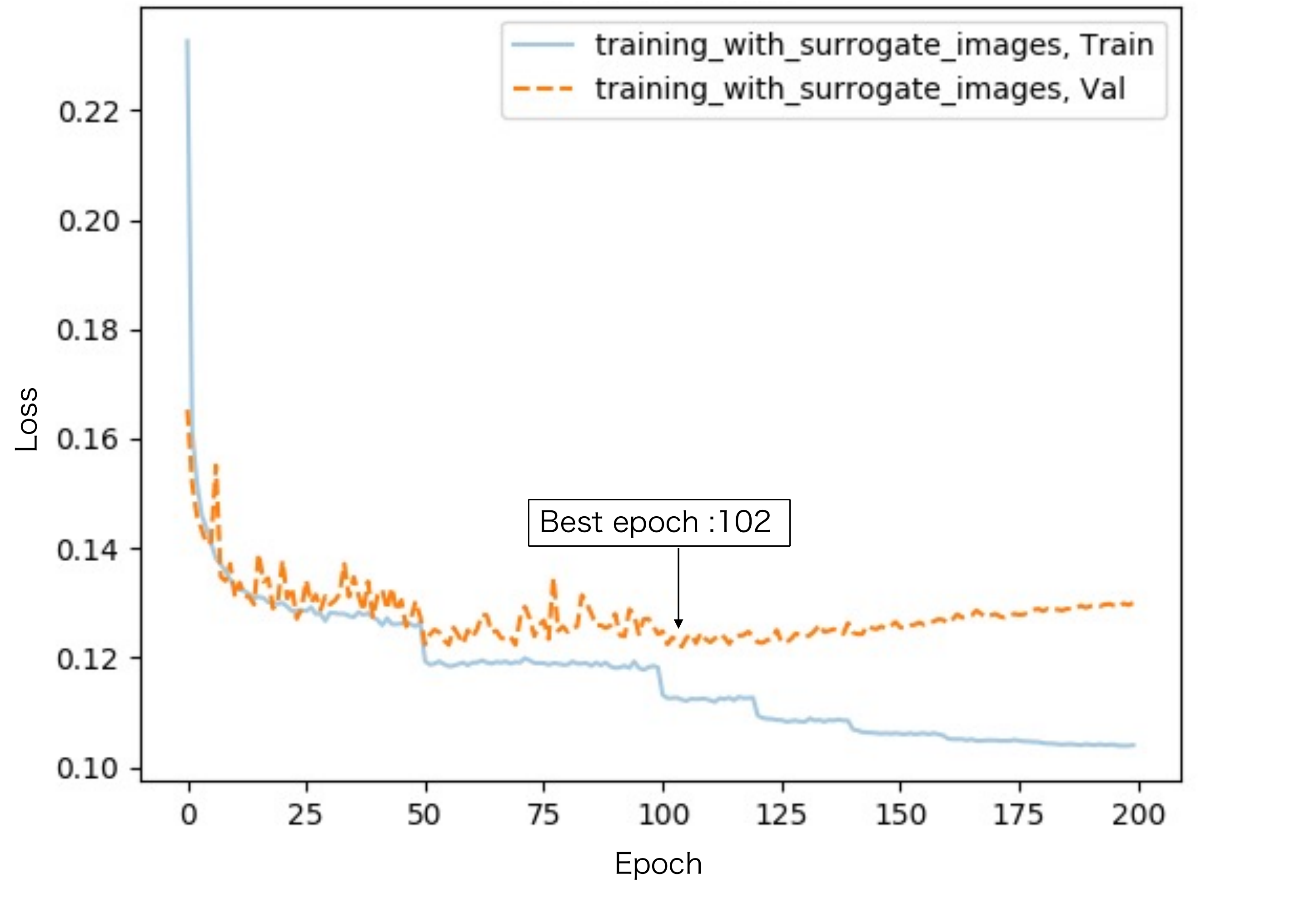}
    \caption{Plots of loss for training and validation data as functions of the number of epochs.
    The solid and dashed lines show the loss for training and validation datasets.
    The epoch with the smallest value in the curve owing to smoothing defined by Eq. (\ref{equ:smoothed_loss}) was determined to be the best epoch, and the model at $\rm{epoch} = 102$ was employed for the subsequent inference.
\label{fig:training_curve}}
\end{figure}

The programs for training and inference were implemented based on PyTorch, and its repository of sample code is publicly accessible \cite{pytorch, pytorch_tutrial}.
The adopted hyperparameters for training are summarized in Table~\ref{tab:params_maskrcnn}.
An amount of 30k images were divided in a ratio of four for training data to one for validation.
The training loss is smoothed using the following equation \cite{Yoshida_cnn}: 
\begin{equation}
    \rm{smoothed~loss} = 0.9*\rm{(previous~smoothed~loss)} + 0.1*\rm{(current~loss) ,}
    \label{equ:smoothed_loss}
\end{equation}
as shown in Figure~\ref{fig:training_curve}.
The solid and dashed lines represent the losses in the training and validation datasets, respectively.
In this figure, epoch 102, which yielded the lowest loss, was defined as the optimal epoch and used for the subsequent inference.

\begin{table}[htbp]
    \centering
    \caption{Hyperparameters for the training of for $\alpha$-decay detection model\label{tab:params_maskrcnn}}
    \begin{tabular}{c|c}
    Backbone&ResNet50 \cite{resnet}\\
    \hline
    Batch size&8\\
    \hline
    Initial learning rate&0.02\\
    \hline
    Learning rate gamma&0.5\\
    \hline
    Learning rate steps&50, 100, 120, 140, 160, 180\\
    \hline
    Momentum&0.9\\
    \hline
    Total epochs&200\\
    \end{tabular}
\end{table}

Before applying the proposed model to real data, its performance was evaluated using a test dataset composed of simulated $\alpha$-decay events generated using the same procedure used for preparing the training and validation datasets.
Three thousand simulated $\alpha$-decay events were prepared and the efficiency of the detection model was evaluated.
In the following discussion, the detection efficiency and purity are calculated as follows:
\begin{equation}
    \rm{Efficiency} = \frac{\rm{(Number~of~detected}~\alpha\rm{\_decay~event)}~}{~\rm{(Number~of}~\alpha\rm{\_decay~events~in~test~dataset)}}
\end{equation}
\begin{equation}
    \rm{Purity}~=\frac{~\rm{(Number~of~detected}~\alpha\rm{\_decay~event)}~}{~\rm{(Number~of~detected~candidates)}}
\end{equation}

\begin{figure}[htbp]
    \centering
    \includegraphics[width=120mm]{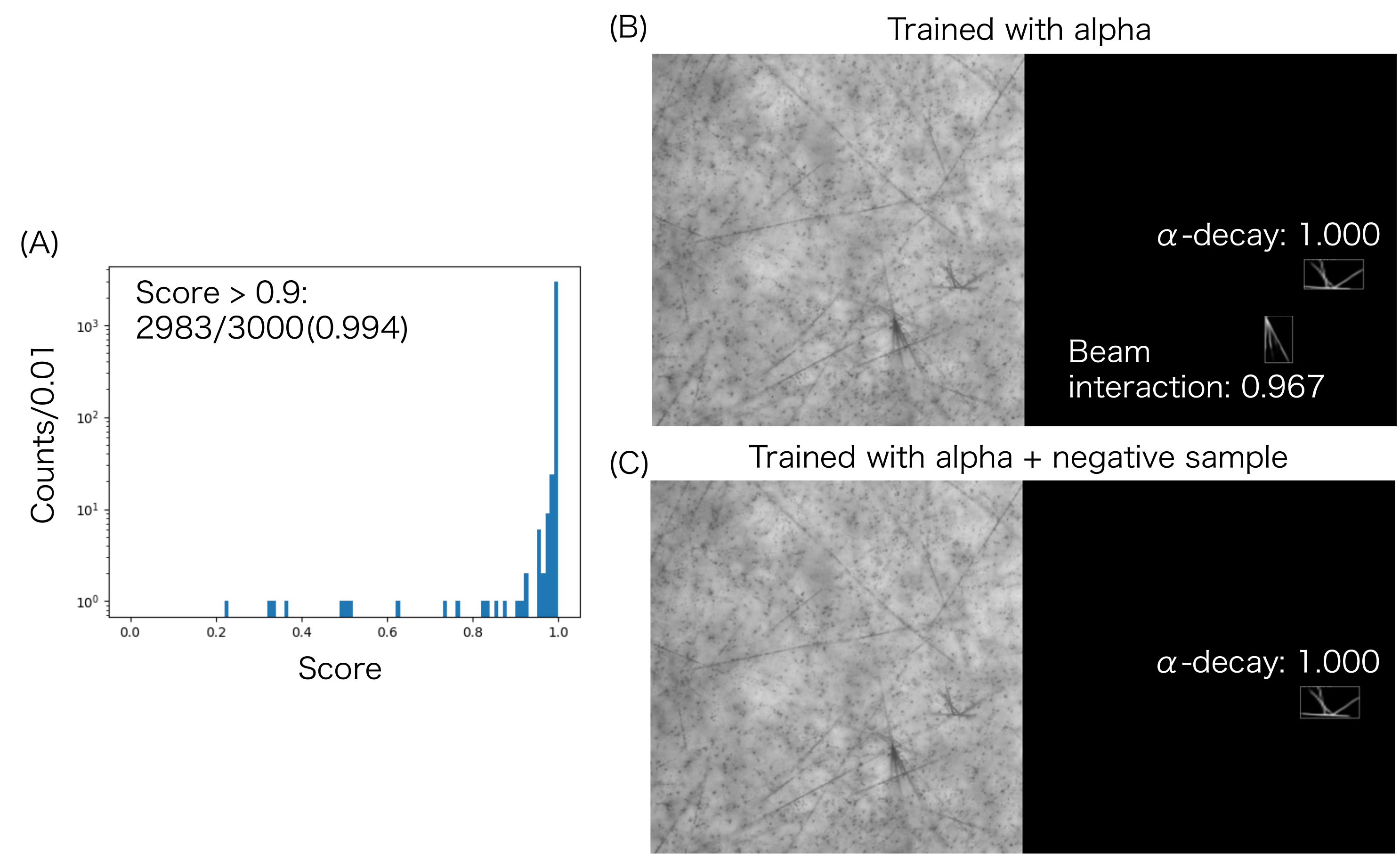}
    \caption{Performance of the developed model for the test image data generated using Monte Carlo simulation and pix2pix.
    Panel (A) shows the score distribution for detected $\alpha$-decay events with images produced with Monte Carlo simulation and pix2pix.
    Three thousand events were analyzed, and $99.4 \pm 0.01$\% of those events were detected with a score over 0.9; therefore, the discrimination threshold on the score for the test with actual emulsion images was set to 0.9.
    Panels (B) and (C) show examples of inference results of models trained with or without negative samples.
    While the beam interaction event was detected with a high score in Panel (B), only the $\alpha$-decay event was correctly detected in Panel (C).
\label{fig:test_surrogated}}
\end{figure}

Panel (A) of Figure~\ref{fig:test_surrogated} shows the score distribution of the objects detected using the developed model.
A score threshold of 0.9 was set to discriminate $\alpha$-decay and others.
Using this criterion, the detection efficiency of the test dataset was 99.4\%.
Panels (B) and (C) in Figure~\ref{fig:test_surrogated} show examples of the differences in the results of the model trained with or without negative samples.
The model trained using only $\alpha$-decay events detected a beam interaction event with a high score of 0.967, as shown in Panel (B).
Using negative samples generated by the JAM simulation, the model learned to eliminate $K^{-}$ beam interaction events and only detect positive samples, as seen in Panel (C).
These results demonstrated the effectiveness of the negative samples included in the training data.

\section{Results and discussions}
We evaluated the performance of the proposed model by applying it to actual images of $\alpha$-decay obtained by the optical scanning of nuclear emulsions.
These were obtained by the visual inspection of 120k microscopic ones in approximately $1.8 ~\rm{cm}^2~\times~0.025~\rm{cm}$ volume of an emulsion sheet.
Over 100 hours were spent on this visual inspection, and 76 $\alpha$-decay events were observed. 

\begin{figure}[htbp]
    \centering
    \includegraphics[width=120mm]{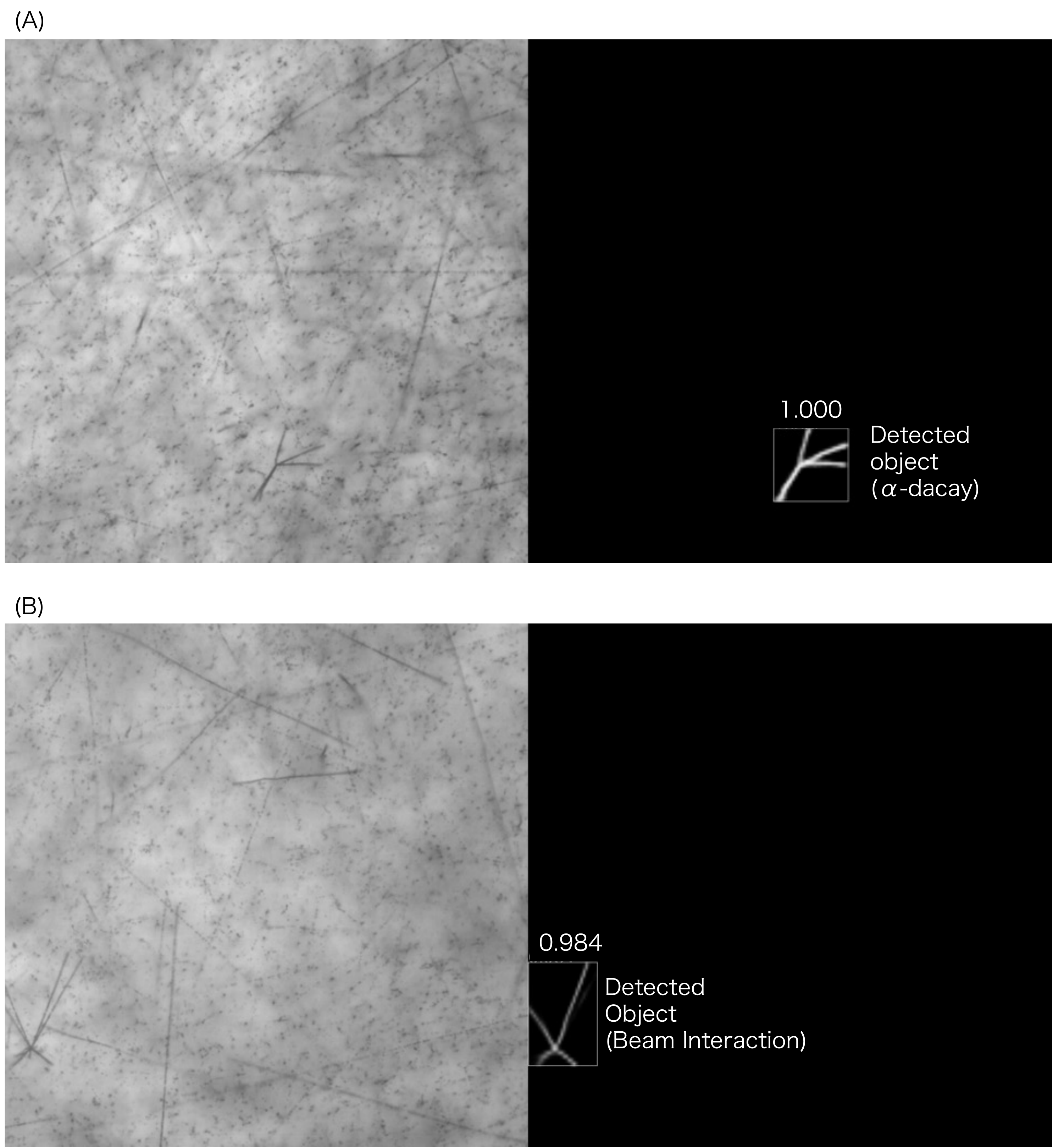}
    \caption{Examples of events detected by the proposed model in real microscopic images. The images on the left side are the input microscope images. The images on the right side are the masks indicating the position and shape of the detected objects. The numbers shown in the figure on the right correspond to the classification score. Panel (A) demonstrates a successful detection of $\alpha$-decay. Panel (B) indicates the detection of a beam interaction event detected as a candidate for $\alpha$-decay. As for the output mask shape, it shares similar characteristics with $\alpha$-decay, but more tracks are observed in the microscopic images than emitted in $\alpha$-decay.
\label{fig:result_real}}
\end{figure}

The performance was evaluated based on whether the developed model properly detected these events.
Figure~\ref{fig:result_real} shows examples of the events detected from real microscopic images using the proposed model.
The images on the left and right sides are microscope images input into the model and masks corresponding to the position and shape of the object detected by the model, respectively.
The numbers shown on the bounding boxes in the figures correspond to the classification scores.
Panel (A) shows a positive result of the $\alpha$-decay detection. 
Panel (B) shows a beam interaction event detected as a candidate for $\alpha$-decay.
The shape of the output mask has characteristics similar to those of $\alpha$-decay, however, the number of tracks observed in the microscopic images is larger than that of typical  $\alpha$-decay events.

The proposed model detected 3572 objects from 120k microscopic images as candidates for $\alpha$-decay events.
To acquire the image data of nuclear emulsions using a microscope, the $x$, $y$, and $z$ axes of the microscope stage were moved by motors to capture varying areas in three dimensions.
The sequence of operations was to move in a direction horizontal to the nuclear emulsion sheet along the $xy$-axis, followed by movement in a direction perpendicular to the sheet along the $z$-axis to change the focal plane.
Events may be counted multiple times as they appear in more than one focal plane during this process.

\begin{figure}[htbp]
    \centering
    \includegraphics[width=120mm]{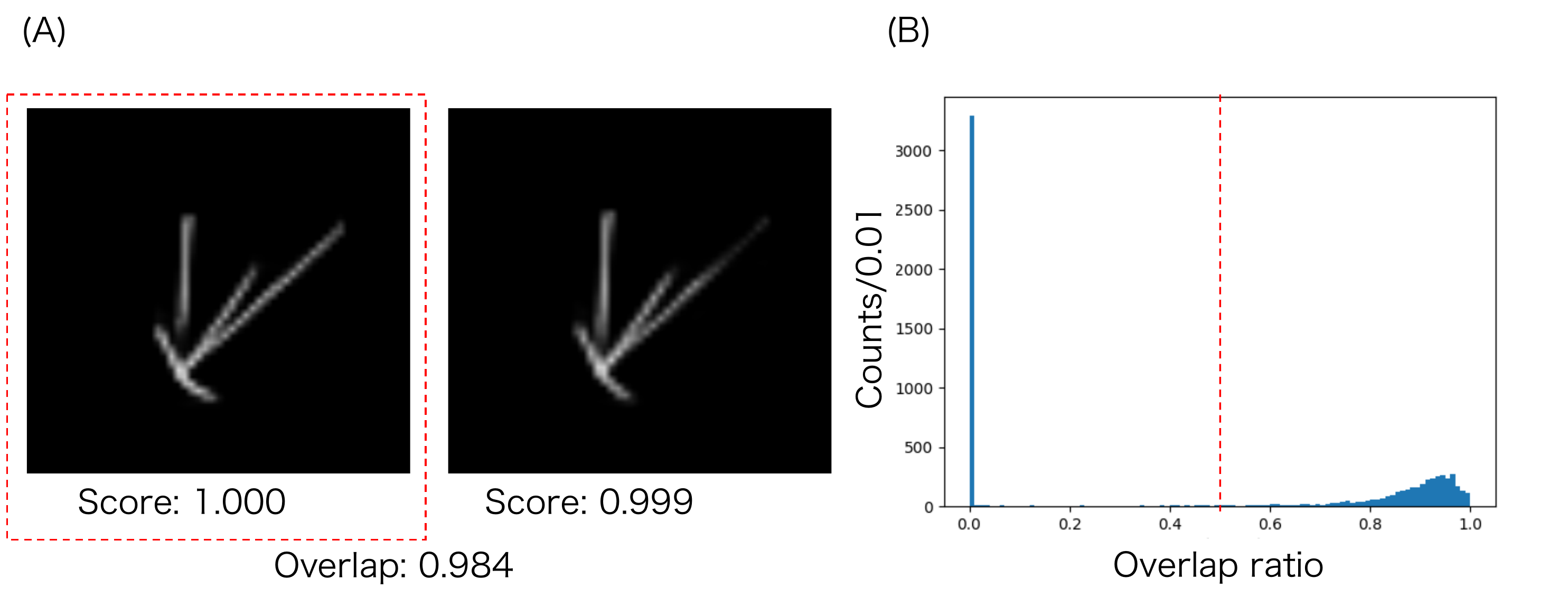}
    \caption{Selection criteria for multiple detected objects using mask information.
    Panel (A) shows the masks for the same $\alpha$-decay event doubly detected in the adjacent depth of field.
    These duplicated events were rejected by calculating the overlap ratio of segmentation results.
    Panel (B) shows the distribution of the overlap ratio for all possible combinations in detected candidates from the same position in the $xy$-plane.
     The dashed line indicates an overlap ratio of less than 0.5 for selected events, with the highest-scoring detected object being chosen among any duplicates.
\label{fig:mask_overlap}}
\end{figure}

To address this issue, duplicate events must be eliminated.
Panel (A) of Figure~\ref{fig:mask_overlap} shows an example of an $\alpha$-decay event detected in this situation.
Because Mask R-CNN also performs segmentation on the detected event to obtain its shape, duplicate events can be eliminated by comparing the shapes of the output masks. 
The mask shapes of the objects shown in the figure achieved an overlap rate of 98.4\%.
The distribution of the overlap ratios of the segmentation results for all possible combinations of events detected at the same position in $xy$-plane is shown in Panel (B).
Events were selected with an overlap ratio of below 0.5 and the object detected with the highest score among the duplicated objects.
Consequently, 2072 unique events remained, and the number of $\alpha$-decay and contaminated background events were evaluated.

\begin{figure}[htbp]
    \centering
    \includegraphics[width=120mm]{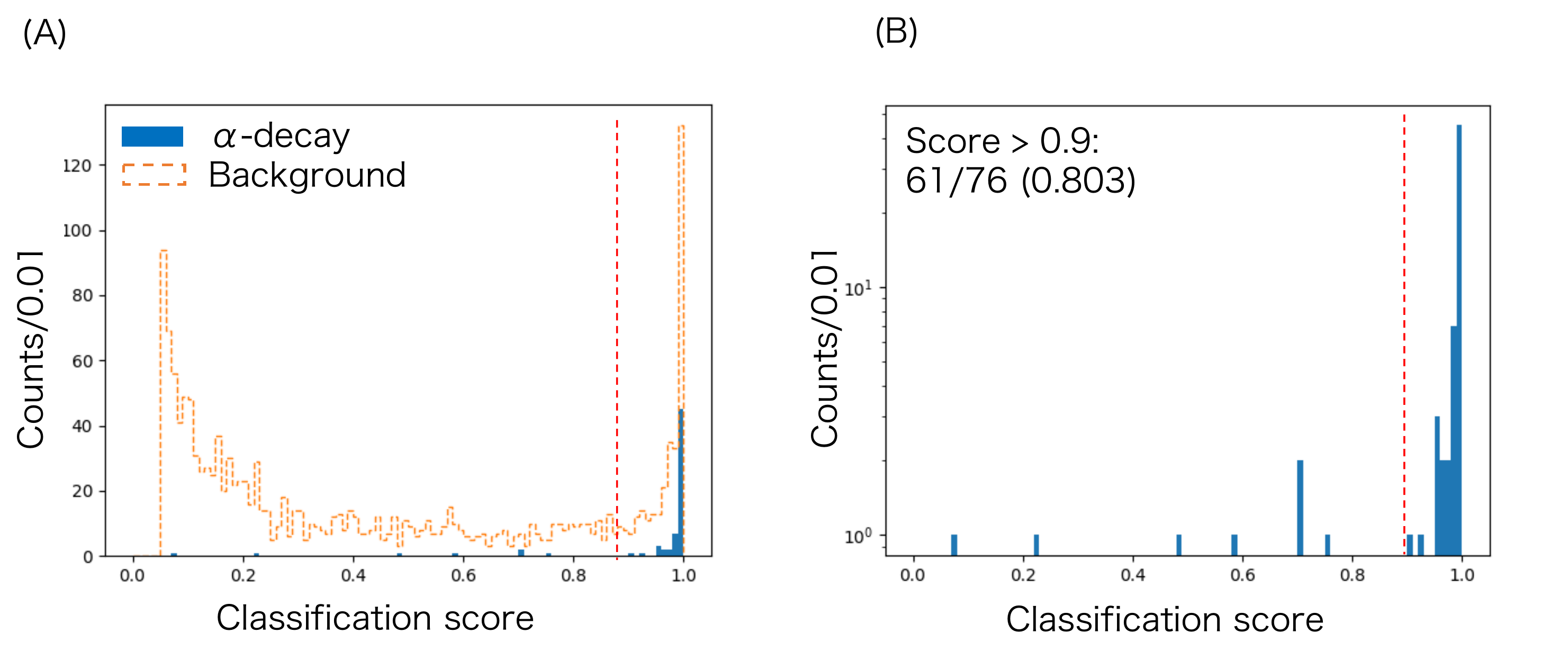}
    \caption{Score distributions of detected objects by the developed model.
    Panel (A) shows the distributions for the number of $\alpha$-decay and other background events represented by the filled histogram and the histogram with the dashed line, respectively.
    Compared to background events, scores for $\alpha$-decay events are narrowly distributed around one.
    Panel (B) shows the entry distribution for $\alpha$-decay event with log-scale, and its shape is comparable to the result for simulated $\alpha$-decay events in the surrogate image as shown in Panel (A) of Figure~\ref{fig:test_surrogated}. 
    The red dashed straight lines in the figures show the score threshold, of 0.9.
\label{fig:score_real}}
\end{figure}

Figure~\ref{fig:score_real} shows the score distributions of the detected objects in nuclear emulsions.
Panel (A) compares the distributions of the $\alpha$-decay events and other background events classified by visual inspection.
The scores for background events were widely distributed, whereas the scores for $\alpha$-decay events were concentrated around one.
Panel (B) shows the distribution for only $\alpha$-decay events on a log scale, and its shape resembles the results for the test data with the surrogate images shown in Panel (A) of Figure~\ref{fig:test_surrogated}.
The red dashed lines in the images indicate a score threshold of 0.9, discussed in Section 3.
By removing events below the threshold, only 482 objects remained out of 2072 objects.
In addition, 61 $\alpha$-decay events were detected among these candidates.
The efficiency and the purity were deduced to be $80.3^{+4.2}_{-4.8}$\% and $12.7^{+0.7}_{-0.8}$\%, respectively.

\begin{figure}[htbp]
    \centering
    \includegraphics[width=120mm]{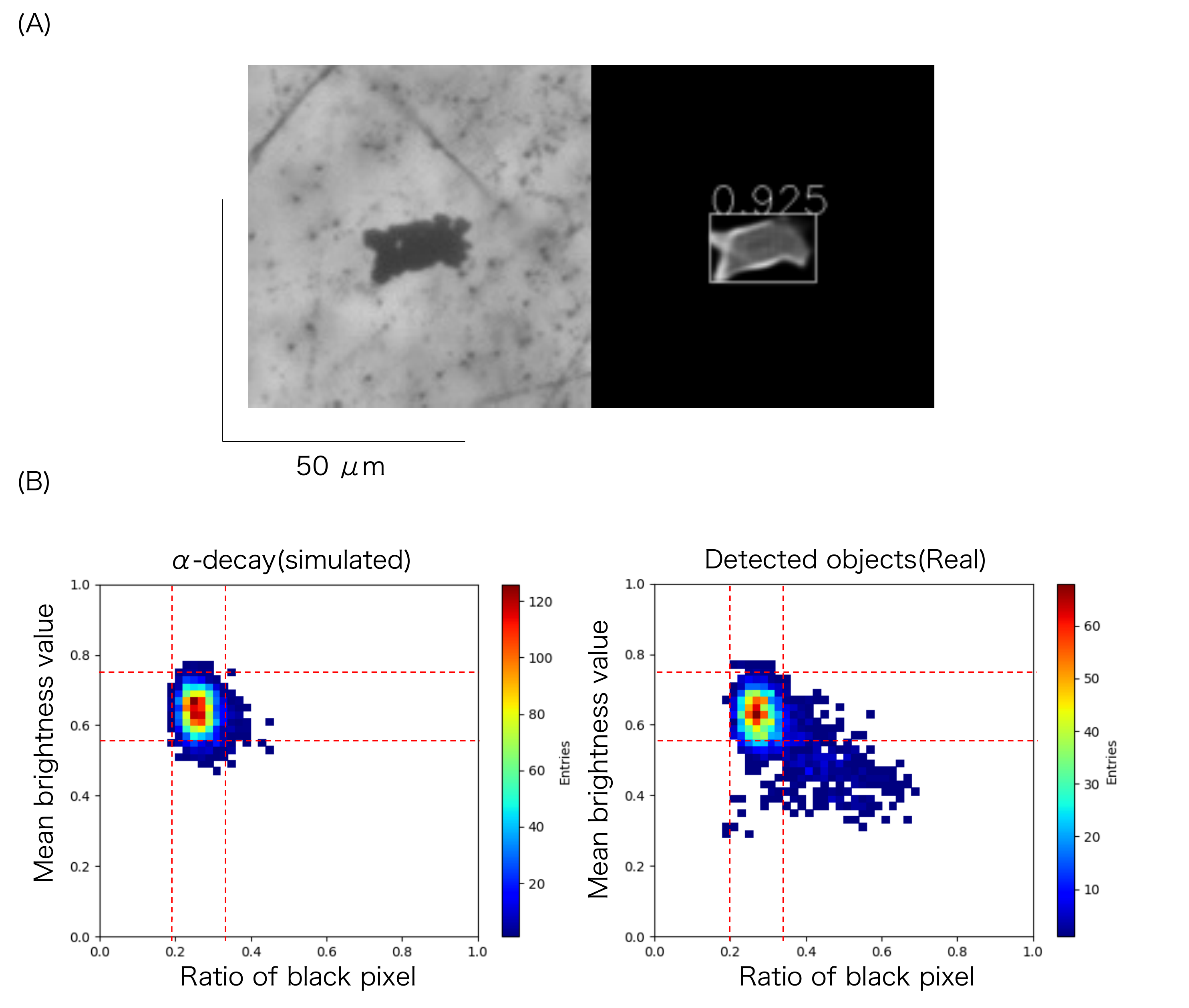}
    \caption{Examples of detected dust events and analysis for reduction.
    Panel (A) shows an example of dust-like objects from shadows created by contaminants during emulsion sheet production. 
    In the developed model, there were several cases where such events were detected with high scores.
    Panel (B) shows two-dimensional correlation histograms with the analyzed results of the mask outputs by the Mask R-CNN-based model.
    The $x$- and $y$-axis of the 2D histogram represent the number of pixels and the average brightness value in the bounding box, respectively.
    The left and right figures show the results for the $\alpha$-decay generated by simulation and for all objects detected in the actual image, respectively.
    Based on the results for the simulated events, only events within the rectangle with red dashed lines in the figure were selected.
\label{fig:dust_reduction}}
\end{figure}

Events detected with scores higher than the threshold were expected to have similar characteristics to $\alpha$-decay events. However, dust-like objects, as shown in sub-panel (c) of Figure~\ref{fig:emulsion_image}, were also detected with high scores.
Panel (A) of Figure~\ref{fig:dust_reduction} shows an example of a dust event detected with a relatively high score of 0.925, although it is completely different in terms of the $\alpha$-decay events.

Therefore, an additional classification method is introduced using the segmentation results.
The discrimination condition was determined by employing the mask information output for the $\alpha$-decay generated in the simulation.
The left figure in Panel (B) of Figure~\ref{fig:dust_reduction} shows the correlation between the number of black pixels in the bounding box and the mean brightness values after binarizing the mask information for $\alpha$-decay events generated by the simulation.
The conditions were determined as areas within the ratio of black pixels from 0.201 to 0.325 and the mean brightness value from 0.547 to 0.731, corresponding to $\pm 2 \sigma$ of the peaks.
When these conditions were applied to the actual images, 352 of the 482 objects remained and 95\% of the dust was eliminated.
Furthermore, none of the 61 $\alpha$-decay events were eliminated, even though the conditions were not adjusted to employ the actual data.
Therefore, the efficiency and the purity were determined as $80.3^{+4.2}_{-4.8}$\% and $17.3^{+0.9}_{-1.0}$\%, respectively.

Finally, the performance of the proposed model was evaluated by comparing it with an analysis of the same test data using the former method.
Event candidates were also detected using the vertex picker, and the detected events were visually classified.
Consequently, 3105 events were selected, including 31 $\alpha$-decay events; that is, the detection efficiency and purity were $40.8^{+5.6}_{-5.5}$\% and $1.0^{+0.1}_{-0.1}$\%, respectively.
Therefore, the proposed Mask R-CNN-based model improved the detection efficiency by a factor of two and the purity by a factor of 12 compared to the former method. 
Furthermore, the filter for eliminating dust-like objects reduced the background events, significantly improving purity.

The performance of the proposed method was also compared with that of a previous trial using a CNN-based classifier \cite{Yoshida_cnn}.
The classifier was trained using real datasets that were manually classified as $\alpha$-decay and other events.
We adopted a classifier for the output images of the vertex picker from the test data used in this study.
The purity improved from 1.0\% to 8.9\% without decreasing the detection efficiency of the vertex picker.

The mask R-CNN-based model achieved an efficiency of $80.3^{+4.2}_{-4.8}$\%, which is twice as high as that of the vertex picker. 
Simultaneously, its purity was 17 times better than that of the vertex picker.
Table~\ref{tab:result_comp_alpha} summarizes the results obtained using each method on the test data.
The model outperformed a CNN classifier trained with real microscopic image data.
Therefore, the effectiveness of the proposed method is confirmed, and the search for events for which no real data exist is realized.

\begin{table}[htbp]
    \centering
    \caption[Comparison of performance of $\alpha$-decay detection and selection]{Comparison of results of $\alpha$-decay detection and selection using previous methods and the developed Mask R-CNN in this work\label{tab:result_comp_alpha}}
    \begin{tabular}{c|c|c|c|c}
    &No. candidates&No. detected&Efficiency [\%]&Purity [\%]\\
    \hline
    \hline
    Vertex picker&3105&\multirow{2}{*}{31/76}&\multirow{2}{*}{$40.8^{+5.6}_{-5.5}$}&$1.0^{+0.1}_{-0.1}$\\
    \cline{1-1} \cline{1-2} \cline{5-5}
    + CNN filter&347&&&$8.9^{+1.1}_{-1.2}$\\
    \hline
    \hline
    Mask R-CNN&482&\multirow{2}{*}{61/76}&\multirow{2}{*}{$80.3^{+4.2}_{-4.8}$}&$12.7^{+0.7}_{-0.8}$\\
    \cline{1-1} \cline{1-2} \cline{5-5}
    + Dust reduction&352&&&$17.3^{+0.9}_{-1.0}$\\
    \hline
    \hline
    \end{tabular}
\end{table}

\section{Conclusion and future work}

We developed a novel method for detecting rare events recorded in nuclear emulsion sheets using machine learning.
The event detection model based on the Mask R-CNN was trained using data obtained by employing Monte Carlo simulations and image style transformations.
The model was applied to real data and its performance was evaluated using $\alpha$-decay events.
The detection efficiency and purity are $80.3^{+4.2}_{-4.8}$\% and $17.3^{+0.9}_{-1.0}$\%, respectively.
Our developed method improves the detection efficiency and purity by a factor of two and 17, respectively, in comparison with the vertex picker based on conventional image processing.
Furthermore, the developed method outperformed the former trial based on a CNN classifier trained using real microscopic images.

This method has been adapted to detect hypernuclear events in real data \cite{Saito_nature}.
The model was trained by generating simulated images of hypernuclear events not identified in the nuclear emulsion data.
The proposed model has successfully detected hypernuclear events \cite{hadron_proc}, and we narrowed candidate events down from as many as two million images to approximately a thousand.
This study enabled the accumulation of many hypernuclear events from nuclear emulsion data within a reasonable period.
 This work opens a new pathway to providing precise experimental data for hypernuclear physics and significantly contributes to understanding nuclear forces
and designing new experiments based on nuclear emulsion technology.

\section*{Acknowledgement}
This work was supported by JSPS KAKENHI Grant Numbers JP16H02180, JP20H00155, JP18H05403, and JP19H05147 (Grant-in-Aid for Scientific Research on Innovative Areas 6005).
The authors thank Risa Kobayashi, Chiho Harisaki, Hanako Kubota and Michi Ando of RIKEN and Yoko Tsuchii of Gifu University for their technical support in mining events in the J-PARC E07 nuclear emulsions.
The authors thank Yukiko Kurakata of RIKEN including the administrative works.



\bibliography{mybibfile}
\end{document}